# Physical properties of transparent perovskite oxides (Ba,La)SnO$_3$ with high electrical mobility at room temperature

Hyung Joon Kim[1], Useong Kim[2], Tai Hoon Kim[1], Jiyeon Kim[2], Hoon Min Kim[2], Byung-Gu Jeon[1], Woong-Jhae Lee[1], Hyo Sik Mun[2], Kwang Taek Hong[1], Jaejun Yu[2], Kookrin Char[2][†] and Kee Hoon Kim[1][†]

[1]*Center for Novel States of Complex Materials Research, Department of Physics and Astronomy, Seoul National University, Seoul 151-747, Republic of Korea*

[2]*Center for Strongly Correlated Materials Research, Department of Physics and Astronomy, Seoul National University, Seoul 151-747, Republic of Korea*

ABSTRACT

Transparent electronic materials are increasingly in demand for a variety of optoelectronic applications, ranging from passive transparent conductive windows to active thin film transistors. BaSnO$_3$ is a semiconducting oxide with a large band gap of more than 3.1 eV. Recently, we discovered that BaSnO$_3$ doped with a few percent of La exhibits unusually high electrical mobility of 320 cm$^2$(Vs)$^{-1}$ at room temperature and superior thermal stability at high temperatures [H. J. Kim *et al*. Appl. Phys. Express. **5**, 061102 (2012)]. Following that work, we report here various physical properties of (Ba,La)SnO$_3$ single crystals and epitaxial films including temperature-dependent transport and phonon properties, optical properties and first-principles calculations. We find that almost doping-independent mobility of 200-300 cm$^2$(Vs)$^{-1}$ is realized in the single crystals in a broad doping range from 1.0×10$^{19}$ to 4.0×10$^{20}$ cm$^{-3}$. Moreover, the conductivity of ~10$^4$ Ω$^{-1}$cm$^{-1}$ reached at the latter carrier density is comparable to the highest value previously reported in the transparent conducting oxides. We attribute the high mobility to several physical properties of (Ba,La)SnO$_3$: a small effective mass coming from the ideal Sn-O-Sn bonding in a cubic perovskite network, small disorder effects due to the doping away from the SnO$_2$ conduction path, and reduced carrier scattering due to the high dielectric constant. The observation of a reduced mobility of ~ 70 cm$^2$(Vs)$^{-1}$ in the epitaxial films is mainly attributed to additional carrier-scattering due to dislocations and grain boundaries, which are presumably created by the lattice mismatch between the substrate $SrTiO_3$ and $(Ba,La)SnO_3$. The main optical gap coming from the charge transfer from O $2p$ to Sn $5s$ bands in $(Ba,La)SnO_3$ single crystals remained at about 3.33 eV and the in-gap states only slightly increased, thus maintaining optical transparency in the visible spectral region. Based on all these results, we suggest that the doped $BaSnO_3$ system holds great potential for realizing all perovskite-based, transparent high-temperature high-power functional devices as well as highly mobile two-dimensional electron gas via interface control of heterostructured films.

## I. INTRODUCTION

Transparent oxide materials offer great potential for a wide range of device applications, e.g. transparent conductive windows for solar cells, flat panel displays, light emitting diodes, and transparent logic devices. Numerous oxide materials including ZnO, $In_2O_3$, and $SnO_2$ have been investigated for these purposes, successfully demonstrating passive conducting leads to active semiconducting devices such as *pn* junctions, field effect transistors, and UV lasers.[1-8] However, these well-known material systems still have limitations so that there is currently active scientific research to find alternative transparent materials which can potentially exhibit better physical properties. For example, high mobility is one essential ingredient for developing transparent logic devices with high density, and oxygen stability is also crucial to overcome interface degradation problems in the oxide *pn* junctions. Moreover, controlling and minimizing the density of defects or dislocations becomes also important to realize useful devices based on transparent conducting oxides (TCOs) and transparent oxide semiconductors (TOSs).

Oxide materials with the perovskite structure have exhibited a plethora of interesting physical properties such as superconductivity, colossal magnetoresistance, ferroelectricity, piezoelectricity, and multiferroicity.[9-12] Extensive research efforts have been made to utilize such varied physical properties in the form of thin film heterostructures.[13] The perovskite materials have also been investigated in the context of searching for new TCOs and TOSs as doped $SrTiO_3$ and doped $CaTiO_3$.[14-18] The alkaline earth stannates with a chemical formula $ASnO_3$ ($A$ = Ba, Sr, and Ca) are another perovskite system that has been used in industry for photoelectrochemical energy conversions, stable capacitors, and gas sensors.[19-21]. In particular, $BaSnO_3$ is known to form an ideal cubic perovskite structure (See, Fig. 1(a)), in which the Sn-O-Sn bonding angle is close to 180°, and corresponds to a transparent, wide band gap semiconductor with an optical gap of ~ 3.1 eV.[22] In order to have electrical conduction in $BaSnO_3$, both Ba and Sn sites were doped to form compounds like $(Ba,La)SnO_3$ and $Ba(Sn,Sb)O_3$ in polycrystalline or thin film forms.[23-27] However, the reported electrical mobility in the thin film was relatively low, less than 2 cm$^2$(Vs)$^{-1}$ and the conductivity reached at most 500 Ω$^{-1}$cm$^{-1}$.[26,27] As those reported transport properties might not be intrinsic, possibly due to, e.g., grain boundary scattering or off-stoichiometry, the growth of high quality single crystals of doped BaSnO$_3$ and the study of their intrinsic physical properties are of high interest.

Recently, we succeeded in growing single crystals[28-29] of (Ba,La)SnO$_3$ and Ba(Sn,Sb)O$_3$ by the Cu$_2$O-flux method up to a lateral size of ~2 mm. In particular, we discovered that a (Ba,La)SnO$_3$ single crystal shows an unprecedentedly high electrical mobility of ~ 320 cm$^2$(Vs)$^{-1}$ with a carrier density of 1.0×10$^{20}$ cm$^{-3}$ at room temperature.[28] In parallel with our discovery, there was a similar report on the growth of (Ba,La)SnO$_3$ single crystals by PbO and PbF$_2$ flux,[30] in which the highest electrical mobility was found to be ~103 cm$^2$(Vs)$^{-1}$. We also found that the epitaxial thin films of (Ba,La)SnO$_3$ grown on SrTiO$_3$ (001) substrate shows an electrical mobility as high as 70 cm$^2$(Vs)$^{-1}$, which is much larger than the previously reported values[26,27] in thin films. Moreover, the resistivity of the (Ba,La)SnO$_3$ film has shown little change even at 500 °C in air, which suggests extremely good oxygen stability and related physical properties at high temperatures.[28]

In order to understand further the origins of such high mobility in the (Ba,La)SnO$_3$ system, we systematically report various physical properties of the single crystals and epitaxial films in the following sections. In Sec. II, we describe the various experimental tools we have employed. In Sec. III, we present experimental and theoretical results in wide doping ranges in the crystals and films, including temperature-dependent transport properties, structural characteristics, band structure calculations, and optical properties. From the Hall effect measurements in a broad doping range from 1.0×10$^{19}$ to 4.0×10$^{20}$ cm$^{-3}$, we found that the mobility remains as high as ~200-300 cm$^2$(Vs)$^{-1}$ and does not show clear dependence on the carrier density. In Sec. IV, we discuss mainly the possible physical origins for obtaining the high electrical mobility. In Sec. V, we conclude that (Ba,La)SnO$_3$ has a great potential for being used as perovskite-based TOSs with high electrical mobility and superior thermal stability.

## II. EXPERIMENTAL

For the single crystal growth of $Ba_{1-x}La_xSnO_3$ ($x$ = 0-0.04), the corresponding polycrystalline specimen was first synthesized as a seed material. High purity $BaCO_3$, $SnO_2$, and $La_2O_3$ powders were weighed in a stoichiometric ratio, thoroughly mixed in an agate mortar, and fired at 1250 °C for 6 hours. After several intermediate grindings, the powders were pressed into a pellet and sintered at 1400-1450 °C for 24-48 hours. For the growth of $(Ba,La)SnO_3$ single crystals, the $Cu_2O$ flux and the sintered powder were placed in a Pt crucible with a molar ratio of about 15:1 and were fired in air above 1250 °C and then slowly cooled down to 1210 °C, followed by a slow furnace cooling to room temperature. The crystals show a cube-like shape as shown in Fig. 1(b).

We grew epitaxial films by use of the pulsed laser deposition technique, using $BaSnO_3$, $Ba_{0.96}La_{0.04}SnO_3$, and $Ba_{0.93}La_{0.07}SnO_3$ targets. We used single crystal $SrTiO_3$ (001) as a substrate, and performed the deposition in an $O_2$ pressure of 100 mTorr at 750 °C. After the deposition, we cooled the samples in an $O_2$ pressure of about 600 mTorr.

We investigated the structural properties of the poly- and single-crystalline samples by use of X-ray diffraction (XRD), for which we used a high power X-ray diffractometer equipped with a single Cu $K_{\alpha 1}$ source (Empyrean$^{TM}$, PANalytical). We performed temperature-dependent X-ray diffraction measurements for the polycrystalline powder of $Ba_{0.999}La_{0.001}SnO_3$ by use of a closed-cycle cryostat (PheniX$^{TM}$, Oxford Instruments), operating in the 20-300 K temperature range. We also characterized the structure of epitaxial films with a four-circle X-ray diffractometer.

The first-principles density functional theory (DFT) calculations were performed by using the Vienna *ab-initio* Simulation Package (VASP) code.[31] The projector-augmented-wave (PAW) method[32] and the Ceperley-Alder parameterization were employed within the local density approximation (LDA).[33] To determine the electronic structure of La-doped $BaSnO_3$, we adopted a 3×3×3 supercell containing 135 atoms and carried out the *k*-space integration using a 2×2×2 mesh within the Monkhorst-Pack *k*-point sampling. For the calculations, all atomic coordinates were relaxed and the Hellmann-Feynman force reached to less than 0.02 eV Å$^{-1}$. The cut-off energy used for the plane wave basis set was 520 eV.

Temperature-dependent resistivity was measured by the conventional four-probe technique for the single crystals and films from 2 K to 300 K either in a closed cycle refrigerator or in a physical property measurement system (PPMS$^{TM}$, Quantum Design). The five-wire configuration was employed to investigate the Hall effect in single crystals and some thin films as a function of temperatures in a magnetic field up to 9 tesla. To make electrical contacts, the single crystals were cut and polished into a rectangular parallelepiped, of which lateral size and thickness were typically ~ 1.5×0.8 mm$^2$ and ~ 50 μm, respectively, as shown in Fig. 1 (b). By use of a shadow mask, we subsequently deposited 100 nm thick Au films as electrical pads on the top of the polished surface. Two electrical pads having a line shape were deposited at both ends of the specimen as current pads. Similarly, four rectangular pads were deposited for the longitudinal (resistivity) and transverse (Hall effect) voltage measurements. In particular, to calculate the transport data accurately, we have tried to minimize the size of the rectangular pads to be less than 100×100 μm$^2$. Furthermore, we attached thin Au wires, of which the diameter is 50 μm, inside the electrical pads with care by use of silver epoxy. Detailed shape for the electrical contacts can be found in the bottom of Fig. 1 (b). A digital ruler was used to obtain accurately the geometry of the sample size and the distance of the electrical leads.

Optical transmission spectra were obtained in BaSnO$_3$ and (Ba,La)SnO$_3$ ($n$ = 2.39×10$^{20}$ cm$^{-3}$) single crystals, which were optically polished to have thickness less than 100 μm. The ultraviolet-visible-near infrared transmission spectra were measured by a fiber-optic spectrometer (StellaNet, EPP2000) equipped with a Xe-arc lamp as a light source. For the absorption coefficient ($\alpha$), we repeated the transmission measurements of the same piece of sample in two different thicknesses and calculated $\alpha$ based on the formula $\alpha$ = -$ln(T_{thick}/T_{thin})/d$, where $d$ is the thickness difference.

Heat capacity data of BaSnO$_3$ single crystals were measured by using a commercial calorimeter provided by PPMS$^{TM}$ (Quantum Design). A custom-made program was developed to perform a least-square-fitting based on the relaxation method with the lumped-$\tau_2$ model,[34] which is often called as the curve fitting method.[35] Based on the fitting, we first extracted the heat capacity of the sapphire platform and thermal grease (addenda) in a wide temperature range from 2 K to 340 K. After obtaining the addenda heat capacity, we subsequently measured the total heat capacity including the BaSnO$_3$ single crystal (total mass = 20.33 mg) and the addenda to extract the heat capacity of the single crystal.

## III. RESULTS

### A. Chemical and structural properties

Figure 1 (c) shows the XRD pattern ($\theta$-$2\theta$ scan) of a typical (Ba,La)SnO$_3$ single crystal after grinding into a fine powder form. The pattern reveals that the single crystal forms the cubic perovskite phase without any detectable impurity phase. We have indeed tried to grow high quality single crystals by using various kinds of fluxes to investigate the intrinsic properties of La-doped BaSnO$_3$. Among those trials, we could also grow rather a large size BaSnO$_3$ single crystals (~2×2×1 mm$^3$) by using the mixed flux of PbO and PbF$_2$, which is similar to the recent report.[30] However, the EPMA (electron-probe-micro-analyses) study[28] showed that the grown single crystals using the Pb-containing flux unavoidably contained significant Pb impurities. On the other hand, the crystals grown by the Cu$_2$O flux did not contain any Cu related impurity according to the EPMA study. It is thus expected that our single crystals without any Cu impurity from the Cu$_2$O flux can offer a better chance to observe the intrinsic properties of (Ba,La)SnO$_3$.

In the case of (Ba,La)SnO$_3$ thin films, the XRD patterns in Fig. 1 (d) show the (00$l$) peaks, due to the epitaxial growth on SrTiO$_3$ (001) as demonstrated in the reciprocal mapping study in Ref. 28. The inset in Fig. 1 (d) shows that the full-width-at-half-maximum (FWHM) in the rocking curve is only 0.09°, supporting a high degree of crystallinity in our epitaxial film. It is worth mentioning that the FWHM in a previous report was about 0.57°,[26] which is much larger than that of our films. This experimental result suggests that the quality of the thin films should be superior to those studied in previous publications.[26,27]

Figure 2 shows the variation of the lattice parameters of polycrystalline $Ba_{1-x}La_xSnO_3$ ($x$ = 0.00, 0.01, 0.02, and 0.04) as obtained from the Rietveld refinement of the powder diffraction data. It is noted that as the La concentration ($x$) increases, the lattice parameter increases almost linearly. It is rather unusual to find such an increase of the lattice parameter while doping by the $La^{3+}$ ion, which has a smaller ionic radius (0.103 nm) than that of $Ba^{2+}$ (0.135 nm).[36] We will discuss the origin of this unusual lattice expansion in a coming section B below.

### B. First-principles calculations

Figure 3 presents the band structures of both $BaSnO_3$ and $(Ba,La)SnO_3$ as obtained by first-principles density functional theory (DFT) calculations with local density approximation (LDA). To our knowledge, the band structure calculation based on the LDA approximation was reported only for $BaSnO_3$,[22] not for $(Ba,La)SnO_3$. The band structures of $BaSnO_3$ and $(Ba,La)SnO_3$ were drawn for the 3×3×3 supercell with 27 unit cells corresponding to the doping rate of $x$ = 0.037 in $Ba_{1-x}La_xSnO_3$. According to the folded Brillouin zone of the 3×3×3 supercell as shown in Fig. 3 (a), $BaSnO_3$ seems to have an indirect band gap. Moreover, based on the band structure in Fig. 3 (b), the La-doping provides electron states directly well-inside the conduction band of $BaSnO_3$, which is mainly characterized as the Sn 5$s$ states with the Sn-O antibonding character. This means that a La ion acts as an electron donor and its ionic valence becomes $La^{3+}$. The occupation in the anti-bonding state is likely to result in repulsive forces between Sn and O to lower the total energy of the crystal structure, thereby inducing expansion of the lattice. The theoretically calculated equilibrium lattice constants of $BaSnO_3$ and 3.7% La doped $BaSnO_3$ are found to be 4.098 Å and 4.102 Å, respectively, showing clearly the increasing trend upon La doping, consistent with experimental results. This theoretical result supports that the experimentally found lattice expansion upon La doping comes from the anti-bonding nature of conduction bands.

It is further noted that the electronic structure of BaSnO$_3$ has a highly dispersive conduction band, mainly composed of Sn 5$s$ bands, mainly due to the presence of an ideal Sn-O-Sn network with its bonding angle close to 180°.[22] The shape of the conduction band is indeed similar to the LDA result of the In$_2$O$_3$[37], a mother compound of the well-known TCO material, In$_2$O$_3$:Sn. This observation indicates that similar to In$_2$O$_3$:Sn, there is a good possibility of obtaining highly mobile $n$-type carriers in BaSnO$_3$ upon doping. When the La doping is made to form (Ba,La)SnO$_3$, the Fermi level is formed well inside the Sn 5$s$ band (the dotted line in Fig. 3 (b)). The change of the band structure by La-doping is negligible near the Fermi level. The presence of La 4$f$ bands at 2 eV above the Fermi level did not affect the band dispersion of the original conduction band of BaSnO$_3$. The calculated effective masses of the conduction band turn out to be about 0.4$m_0$ for both BaSnO$_3$ and (Ba,La)SnO$_3$, where $m_0$ is the free electron mass. The Fermi level of (Ba,La)SnO$_3$ in Fig. 3 (b) indicates that the 3.7 % La doped (Ba,La)SnO$_3$ is in a degenerately doped regime with $n$-type carriers, which is consistent with the temperature-dependent transport results explained below.

C. Transport properties

Figure 4 summarizes the resistivity and mobility variation as a function of carrier density at room temperature in both single crystals and thin films. The data from our earlier report are also compared (closed squares and triangles). Previously,[28] we found that the mobility of single crystals is roughly proportional to $n^{-1}$, while the mobility of thin films is proportional to $n^{1/2}$ in a low doping range smaller than $n = 4\times10^{20}$ cm$^{-3}$. The $n^{-1}$ dependence of single crystals could be understood as an ionized-dopant scattering effect, which is known to be roughly proportional to the number of scattering centers, i.e., ionized dopant density.[28] In the case of films, the $n^{1/2}$ dependence could be mainly attributed to the dominant dislocation or grain boundary scattering effects, which are increasingly screened in roughly proportional to the square root of the mobile carrier density. The existence of the dislocations or grain boundaries in the films is expected due to the large lattice mismatch between (Ba,La)SnO$_3$ and SrTiO$_3$.

On the other hand, in the extended measurements for the single crystals in a broad doping range from $1.0 \times 10^{19}$ to $4.0 \times 10^{20}$ cm$^{-3}$, we find that the measured electrical mobility values are instead scattered between 200 - 300 cm$^2$(Vs)$^{-1}$ without showing clear dependence on the carrier density. It is noteworthy that the previous data for the single crystals (solid squares in Fig 4.) show rather large variation of mobility and reduced mobility at a high carrier density although the crystals come from the same batch. Thus, we tentatively presume that the crystals studied in Ref. 28 might have systematic increase of oxygen vacancy with carrier density increase. We note that the present newly updated mobility values still mark the highest record among the wide-band-gap oxide semiconductors in a similar doping level; e.g., highest mobility around the doping level of $n = \sim 1\times10^{20}$ cm$^{-3}$ was ~160 cm$^2$(Vs)$^{-1}$ for indium oxide (In$_2$O$_3$),[38] ~ 50 cm$^2$(Vs)$^{-1}$ for tin oxide (SnO$_2$),[39] and ~100 cm$^2$(Vs)$^{-1}$ for zinc oxide (ZnO).[40] We further note that, due to the high mobility of the (Ba,La)SnO$_3$ system, the conductivity value in the high doping regions above $n = 2\times10^{20}$ cm$^{-3}$ is ~10$^4$ $\Omega^{-1}$cm$^{-1}$ in the single crystal at room temperature (Fig. 4 (a)), which is close to the highest value among the transparent conductors. The best conductivity of the In$_2$O$_3$:Sn is known to be ~10$^4$ $\Omega^{-1}$cm$^{-1}$. Thus, (Ba,La)SnO$_3$ can be also used as transparent conductors.

To understand further detailed electrical transport properties, we performed the temperature-dependent resistivity measurements as summarized in Fig. 5. First, we note that the carrier densities are almost temperature-independent in both single crystals and films (Figs. 5 (a) and (b)). The resistivities of both single crystals and films mostly increase as increasing temperature (Figs. 5 (e) and (f)). The positive temperature coefficient of the resistivity and temperature-independent carrier density clearly show the metallic behavior in both films and single crystals, indicating that all the materials are in a degenerately doped semiconducting regime. The resistivity values of the single crystals decrease nearly by a factor of two upon being cooled from room temperature to 2 K (Fig. 5 (e)), which then results in the increase of the corresponding mobility by the same factor (Fig. 5 (c)). The mobility increases at low temperatures seem to be a common property of both crystals and films. The factor of the resistivity decrease from room temperature to 2 K, as observed in the films with $x$ = 0.04 and 0.07, is also reflected in the increase in the corresponding mobility (Fig. 5 (d)). On the other hand, it is found that the residual resistivity values of the films are generally larger than those of the crystals (Fig. 5 (f)) upon being compared at a similar carrier concentration (e.g. one order higher near $n \approx 1 \times 10^{20}$ cm$^{-3}$), implying that there exist extra-scattering sources in the films such as grain boundaries and dislocations.

### D. Optical properties

Figure 6 compares the transmission and optical absorption spectra for the two single crystals, undoped BaSnO$_3$ and (Ba,La)SnO$_3$ with $n$ = 2.39×10$^{20}$ cm$^{-3}$. Transmittance of the BaSnO$_3$ single crystal reaches as high as 0.71 in the visible spectral region (1.8-3.1 eV) although its thickness of 68.7 µm is rather high. The significant suppression of the transmittance as well as a sharp increase in $\alpha$ near 3.1 eV reflects the development of an optical gap. Transmittance of the La doped crystal is progressively suppressed at a low frequency region, resulting in the Drude-type absorption tail, presumpably due to the increase of free carriers. However, the transmittance level in the doped crystal remains finite around 0.2 albeit it is rather thick (~53.7 µm) and has rather a high doping level ($n$ = 2.39×10$^{20}$ cm$^{-3}$). The resultant $\alpha$ spectra of the doped crystal is still less than 300 cm$^{-1}$ in the visible spectral region, predicting that the doped sample should have transmittance of ~ 0.8 in e.g., a thin film having thickness about 100 nm. This high transparency has been also observed in previous optical studies on the thin films grown on SrTiO$_3$ substrates[26]. It is also noted that the steeply increasing part in α has slightly shifted to a higher energy in the doped crystal, indicating the presence of the Burstein-Moss shift[41,42] as often observed in degenerate semiconductors.

According to the band calculation in Fig. 3, $BaSnO_3$ is an indirect gap semiconductor, in which the conduction band minimum and the valence band maximum have predominant Sn 5$s$ and O 2$p$ characters, respectively. To determine experimentally the gap nature of $BaSnO_3$, we plotted the curves of $\alpha^{0.5}$ and $\alpha^2$ vs photon energy for both undoped $BaSnO_3$ and $(Ba,La)SnO_3$ in the inset of Fig. 6 (b). For $BaSnO_3$, we could find a linearly increasing region in both curves. Upon applying a linear extrapolation, the indirect and direct band gaps are estimated as 2.95 eV and 3.10 eV, respectively. This observation suggests that the optical transition with the lowest energy reflects the indirect transition nature while the direct optical transition with a bit higher energy is quite strong in this material. This is somewhat expected because the direct optical transition between the Γ-points of the valence and conduction band satisfies an optical selection rule (from O 2$p$ to Sn 5$s$), while the indirect transition from the valence band maximum (R-point) to the conduction band minimum (Γ-point) are generally low. It is noteworthy that quite similar optical transition characters have been also observed in another cubic perovskite $SrGeO_3$, a new TCO candidate compound reported recently.[43]

The experimentally estimated direct or indirect band gaps (2.95-3.10 eV) in $BaSnO_3$ is much larger than that (~1 eV) estimated from the band structure in Fig. 3. It is rather well-known that the gap predicted by the band calculation based on the LDA approximation underestimates the actual gap in many transparent oxide materials because the LDA calculation doesn't take into account the electron correlation effect properly. Even the well-known TCO, $In_2O_3$:Sn exhibits a similar kind of discrepancy between the LDA calculation and the experimentally determined gap due to the correlation effect[37]. In addition, $In_2O_3$:Sn was found to show yet another discrepancy of 0.81 eV between the experimental optical gap and the gap measured by the photoemission and X-ray emission spectroscopy due to the presence of the optical selection rule.[44] Therefore, it will be interesting to study in future the origin of such an optical gap enhancement in $BaSnO_3$ as well. In reference to the case of In$_2$O$_3$:Sn, it will be worthwhile to check whether the electron correlation effects alone can explain the discrepancy or an additional optical selection becomes necessary.

### E. Debye temperatures

To extract the information on the Debye temperature, we used two methods, i.e., heat capacity and thermal expansion measurements. Figure 7 (a) shows the specific heat of a BaSnO$_3$ single crystal at temperatures between 2 and 330 K. For the heat capacity analysis, we found that the two Debye phonon modes become necessary because one phonon mode alone could not fully explain the experimental data. Based on the Debye theory, the temperature dependent specific heat coming from the two phonon contributions can be described as follows:[45]

$$C_v = 9qNk_B \left[ x(\frac{T}{\Theta_{\text{low}}})^3 \int_0^{\Theta_{low}/T} \frac{x^4 e^x}{(e^x-1)^2} dx + y(\frac{T}{\Theta_{\text{high}}})^3 \int_0^{\Theta_{high}/T} \frac{x^4 e^x}{(e^x-1)^2} dx \right] \quad (1)$$

, where $\Theta_{\text{low}}$ and $\Theta_{\text{high}}$ are the Debye temperatures of each phonon mode, and $x$ and $y$ are the weights of the phonon modes. Open circles represent the temperature-dependent specific heat data of the BaSnO$_3$ single crystal while the lines refer to the two phonon mode contributions (dotted lines) and their sum (solid line). The model fitting can successfully explain the experimental data at least above 20 K and the resultant two Debye temperatures $\Theta_{\text{low}}$ and $\Theta_{\text{high}}$ are found to be 307 K and 950 K, respectively. The resultant $x/y$ ratio is around 1.5, implying that the heat capacity below 300 K is mainly governed by the low energy Debye phonon.

Figure 7 (b) displays the evolution of temperature-dependent lattice constant of a Ba$_{0.999}$La$_{0.001}$SnO$_3$ polycrystalline specimen. Upon increasing the temperature, the lattice parameter increases, consistent with lattice expansion due to anharmonic thermal vibration. To model the thermal expansion data of Ba$_{0.999}$La$_{0.001}$SnO$_3$, we assumed that the two Debye temperatures as obtained from the heat capacity data can explain the intrinsic energy of the lattice and thus its temperature-dependent expansion as well. Then, the temperature dependencies of the lattice parameter ($a$) can be described as:[46]

$$a = a_0 + bT(\frac{T}{\Theta_{\text{low}}})^3 \int_0^{\Theta_{low}/T} \frac{x^3}{e^x-1} dx + cT(\frac{T}{\Theta_{\text{high}}})^3 \int_0^{\Theta_{high}/T} \frac{x^3}{e^x-1} dx \quad (2)$$

, where $a_0$ and $\Theta_D$ are the lattice constant at 0 K and Debye temperature, respectively, and $b$ and $c$ are the constant prefactors. Upon using the same Debye temperatures from heat capacity, we could get satisfactory fitting result, confirming that the two Debye temperatures of 307 K and 950 K are quite effective in describing the lattice energy and related thermal properties. As the two Debye temperatures are quite different, one phonon mode around 300 K seems to be approximately effective in describing the phonon properties of BaSnO$_3$ at low temperatures. For example, upon fitting the thermal expansion data with one Debye temperature of 365 K, we could obtain almost the same good fit results as the two phonon approximation (not shown), supporting that the lower energy phonon mode contributes mostly the thermal expansion as well.

## IV. DISCUSSION

The electrical mobility ($\mu$) in a simple one band model is expressed by an electron effective mass ($m^*$) and a total electron scattering rate ($\tau^{-1}$):

$$\mu = e\tau/m*, \tag{3}$$

where $e$ is an electron charge. Therefore, a high mobility can be realized with a small electron effective mass and a small total electron scattering rate. In the followings, we discuss the origins for achieving such a high mobility in (Ba,La)SnO$_3$ based on the peculiar characteristics reflected in the two key physical parameters, i.e., electron scattering rate and effective mass.

### A. High mobility in the crystals and thin films

Controlling impurity scattering seems to be important to obtain the highest mobility of ~ 320 cm$^2$(Vs)$^{-1}$ in (Ba,La)SnO$_3$ single crystals as observed in our previous study[28] and 200- 300 cm$^2$(Vs)$^{-1}$ as found in this work. It was recently reported that single crystals grown by PbO + PbF$_2$ flux produced a mobility[30] close to 103 cm$^2$(Vs)$^{-1}$ at a doping level ~ 8 – 10×10$^{19}$ cm$^{-3}$. As the Pb impurity can enter both Ba and Sn sites in this flux, the observation of lower mobility implies that the presence of the additional impurity other than La might lead to the increased dopant scattering. As mentioned in Sec. III A, the absence of Cu impurity in our single crystals should be then helpful in reducing the additional impurity scattering and obtaining physical properties close to the intrinsic ones.

In the thin films, it is interesting to note that a previous study found the mobility to be at best 0.69 $cm^2(Vs)^{-1}$ although the reported (Ba,La)SnO$_3$ films were epitaxially grown on the same type of substrate, SrTiO$_3$ (001),[26] as the present work. The FWHM in the rocking curve was 0.57° in the previous study[26] while it is 0.09° in our study (Fig. 1 (d)), which represents a conspicuous improvement in the crystallinity. Therefore, the highest electrical mobility of ~ 70 $cm^2(Vs)^{-1}$ realized in our thin films is most likely to be associated with the superior structural properties.

On the other hand, the highest mobility of ~ 70 $cm^2(Vs)^{-1}$ in our films is still much lower than the values in the single crystals. Even though our films were epitaxially grown, we found from the transmission electron microscopy study that there exist significant grain boundaries and threading dislocations, due to the large lattice mismatch (more than 4%) between the substrate and (Ba,La)SnO$_3$. Those grain boundaries/dislocations are expected to act as double-Schotttky barriers for the electron transport,[47,48] thereby giving rise to lower electrical mobility and higher residual resistivity than single crystals as demonstrated in Fig. 5. Therefore, it is expected that the mobility in thin films will be further improved when the dislocation density gets reduced by, e.g., using BaSnO$_3$ single crystal substrates.

Related to the presence of such grain boundaries/dislocations, the room temperature carrier densities in the films with nominal doping concentration $x$ = 0.01, 0.04, and 0.07 are 7.0×10$^{19}$ cm$^{-3}$, 4.0×10$^{20}$ cm$^{-3}$, and 6.8×10$^{20}$ cm$^{-3}$, respectively. The activation rates of dopants then become 45, 69, and 67 %, respectively. The lower activation rate for the lowest doping indicates existence of the enhanced charge trapping mechanism. Moreover, as evident in Fig. 5 (d), the mobility is reduced significantly when carrier density is decreased.[28] This observation implies that the higher carrier density effectively reduces the effect of dislocation/grain boundary scattering. Once the double Schottky barrier is formed in the films, the higher carrier density is likely to induce thinner barrier width and smear out the barriers, resulting in the enhanced mobility as the carrier density increases up to around $n \sim 4\times10^{20}$ cm$^{-3}$. However, as the doping rate becomes higher than $4\times10^{20}$ cm$^{-3}$, the scattering by ionized dopants seems to start dominating rather than dislocations or grain boundaries, to induce a mobility decrease by the increased dopant level.

## B. Effective mass

To understand the high mobility observed in the (Ba,La)SnO$_3$ single crystals within the band structure framework, one should expect quite a small effective mass. As mentioned in the LDA calculation results (Sec. III B), the effective masses of BaSnO$_3$ and (Ba,La)SnO$_3$ are predicted to be $\sim 0.4m_0$. It turns out that the value of $\sim 0.4m_0$ is comparable to, but not particularly smaller than the theoretically predicted effective masses of other wide-band-gap oxide semiconductors, e.g., In$_2$O$_3$ (0.30$m_0$),[49] SnO$_2$ (0.38$m_0$),[50] and ZnO (0.24$m_0$).[51] The effective mass can be also estimated by the Burstein-Moss shift,[41,42] which predicts the energy difference in the optical band gap ($\Delta E$) between undoped and doped semiconductors as $\Delta E = h^2(3n/\pi)^{2/3}/(8m^*)$, where $h$ is Plank constant and $m^*$ is effective mass. As we estimate $\Delta E$ from the shift in the direct optical gap in the inset of Fig. 6 b, $\Delta E = \sim 0.23$ eV so that the effective mass is estimated as $\sim 0.60m_0$. This value is slightly larger than the theoretical prediction but small enough to guarantee a high mobility. On the other hand, there exist another recent predictions of $m^* = 0.06m_0$[52] or $0.2m_0$[53] based on the LDA calculations based on different approximations such as GGA (generalized gradient approximation) or use of a specific hybid-functional, respectively. The predictions of the effective mass at this stage are thus scattered, ranging from 0.06 to 0.5$m_0$, indicating that it is rather sensitive to the approximation in the DFT calculation. Therefore, to pin down whether the effective mass is a dominant quantity for creating the high mobility or not, it would be necessary to determine an effective mass of (Ba,La)SnO$_3$ more accurately.

C. Effects of various scattering in the crystal

Another key physical quantity to determine the electrical mobility is the electron scattering rate, which can be expressed by a sum of several scattering rates, according to Matthiessen's rule:

$$\tau^{-1} = \sum \tau_i^{-1} \qquad (4)$$

, where $\tau_i^{-1}$ are the electron scattering rates from different scattering sources. In the single crystals of (Ba,La)SnO$_3$ with minimized extrinsic scattering sources, e.g., defects, the electron-phonon and ionized-dopant scatterings are likely to be main sources of the electron scattering. It is generally known that the ionized-dopant scattering in the degenerately doped regime is almost temperature independent. Then, the majority of the temperature-dependent scattering can be attributed to the electron-phonon scattering in the (Ba,La)SnO$_3$ system, in which the degenerately doped regime is clearly realized when the doping rate is higher than $1.0 \times 10^{19}$ cm$^{-3}$. For example, the temperature-dependent resistivity variation should be then determined by the electron-phonon scattering as the carrier density is mostly temperature-independent (Fig. 4 (b)).

The Debye temperature should be closely related to the electron-phonon scattering strength, as the electron-phonon scattering will increase with the number of thermally activated phonons; e.g., a smaller Debye temperature will result in a higher amount of phonon population at a given temperature. The Debye temperatures of (Ba,La)SnO$_3$ are 307 K and 950 K as determined by both heat capacity and thermal expansion measurements. The Debye temperature of 307 K is indeed a bit lower than the known Debye temperatures of In$_2$O$_3$ (420 K),[54] SnO$_2$ (500 K),[54] and, ZnO (399.5 K).[55] Therefore, it is expected that the electron-phonon scattering in (Ba,La)SnO$_3$ should not be particularly smaller to cause the higher mobility than those of the other transparent electronic materials.

Another important contribution to the total scattering rate is the ionized-dopant scattering. In (Ba,La)SnO$_3$, the La$^{3+}$ dopants are the main ionized impurities. In a degenerately doped semiconductor, the mobility due to the ionized impurity ($\mu_{ii}$) can be expressed as:[56,57]

$$\mu_{\text{ii}} = \frac{3(\varepsilon_r\varepsilon_0)^2 h^3}{Z^2 m^{*2} e^3} \frac{n}{N_{\text{i}}} \frac{1}{F_{\text{ii}}(\xi_d)} \quad \text{and} \quad \xi_d = (3\pi^2)^{1/3} \frac{\varepsilon_r\varepsilon_0 h^2 n^{1/3}}{m^* e^2} \tag{5}$$

, where the screening function $F_{\text{ii}}(\xi_d)$ is given by:

$$F_{\text{ii}}(\xi_d) = \left[1 + \frac{4\xi_{np}}{\xi_d}\left(1 - \frac{\xi_{np}}{8}\right)\right]\ln(1+\xi_d) - \frac{\xi_d}{1+\xi_d} - 2\xi_{np}\left(1 - \frac{5\xi_{np}}{16}\right) \quad \text{and} \quad \xi_{np} = 1 - \frac{m_0^*}{m^*}. \tag{6}$$

Here, $h$ is Planck's constant, $N_{\text{i}}$ is a number of ionized dopant, and $\varepsilon_r$ is a dielectric constant. $m_0^*$ is the effective mass at the conduction band edge and $m^* = m_0^*[1 + 2\alpha(E - E_C)]$, where $\alpha$ is the non-parabolicity parameter and $E - E_C$ is the electron energy relative to the conduction band edge ($E_C$). Equation (5) predicts that in the ionized impurity scattering regime, $\mu_{\text{ii}}$ is generally proportional to the square of the dielectric constant. We should note that the dielectric constant of $BaSnO_3$ (about 20)[58] is almost two times larger than the well-known transparent electronic materials; $In_2O_3$ (about 9), $SnO_2$ (9.6 - 13.5), and ZnO (8.75 - 7.8).[56] Therefore, the high mobility state in (Ba,La)$SnO_3$ may be due to greatly reduced ionized-dopant scattering, coming from the enhanced screening strength associated with the high dielectric constant.

Another important factor to reduce the ionized-dopant scattering could be the capability of doping the La dopant into the Ba site. It is expected that the ionized-dopant scattering should be greatly reduced if the dopant is located away from the $SnO_2$ network that are the main conduction paths. This would be favorable for realizing almost defect-free $SnO_2$ conduction paths. Empirical comparison of the residual resistivity supports this reasoning. The residual resistivity of the Ta doped $SnO_2$ was ~ 0.83 m$\Omega$ cm, at $n$ = ~ $1\times10^{20}$ cm$^{-3}$,[39] which is seven times larger than that of (Ba,La)$SnO_3$ single crystal, ~ 0.12 m$\Omega$ cm at $n$ = $1.18\times10^{20}$ cm$^{-3}$ (Fig. 5). Thus, it is postulated that the location of La dopant away from the $SnO_2$ conduction paths plays an important role to minimize the ionized-dopant scattering as well.

One caveat in applying equation (5) to understand the ionized-dopant scattering in our single crystals lies in the screening function $F_{\text{ii}}(\xi_d)$. Once the ionized impurity level ($N_{\text{i}}$) and actual carrier density ($n$) are proportional to each other, $\mu_{\text{ii}}$ is inversely proportional to $F_{\text{ii}}(\xi_d)$. In general, $F_{\text{ii}}(\xi_d)$ increases as the carrier density increases so that one should expect a reduced $\mu_{ii}$ with increase of $n$ in the ionized-dopant scattering regime. However, the new result in Fig. 4 (b) is not yet decisive to draw a conclusion whether the mobility indeed shows any dependence on $n$. The mobility values near the same carrier density level exhibit large scatter. This indicates that the single crystals are not yet perfect so that they might have additional scattering sources such as oxygen vacancies. If the oxygen vacancies are present in the middle of the $SnO_2$ conduction paths, they are expected to act as stronger scattering centers than the $La^{3+}$ ions. Moreover, the actual carrier density can be also fluctuating inside a single crystal or over different pieces, possibly resulting in the scattered mobility and carrier density as seen in Fig. 4 (b). In light of the very small oxygen diffusion constant in $BaSnO_3$ material system, as we reported previously,[28] full and uniform oxygenation of mm-thick crystals may be difficult. Further studies are underway to reduce such additional scattering sources.

## V.    CONCLUSIONS

We found a high electrical mobility in the transparent perovskite material, $(Ba,La)SnO_3$. The $(Ba,La)SnO_3$ single crystals grown by the $Cu_2O$ flux method showed an electrical mobility of 200-300 $cm^2(Vs)^{-1}$ in a broad doping range from $1.0\times10^{19}$ to $4.0\times10^{20}$ $cm^{-3}$, constituting the highest value among the known transparent electronic materials, such as doped $In_2O_3$, $SnO_2$, and $ZnO$. In the thin films grown epitaxially on $SrTiO_3$ (001), the maximum mobility reached only ~ 70 $cm^2(Vs)^{-1}$ due to the presence of dislocations/grain boundaries, which points out the possibility of enhancing the mobility in the thin film form upon having a proper substrate material. We discussed that the high mobility can arise from several unique physical conditions of $(Ba,La)SnO_3$: highly dispersive Sn $5s$ band coming from an ideal Sn-O-Sn bonding angle close to 180° and minimized ionized-dopant scattering from the high dielectric constant and the capability of putting $La^{3+}$ dopant away from the main $SnO_2$ conduction paths. Moreover, it is noteworthy that such a high mobility is linked to the high conductivity ~ $10^4$ $\Omega^{-1}cm^{-1}$ at room temperature at a doping level of $n = 2\times10^{20}$ $cm^{-3}$, which is comparable to the highest value among the transparent conducting oxides. Therefore, combined with the thermal stability of the oxygen diffusion, the (Ba,La)SnO$_3$ system offers an unprecedented opportunity for realizing all perovskite based, transparent high-temperature, high-power functional devices.

## ACKNOWLEDGEMENTS

We thank Tae Won Noh, Yun Daniel Park, and Tak Hee Lee for discussions. We acknowledge support from the NRF through Accelerated Research Program (R17-2008-33-01000-0) and through the Creative Research Initiative (2010-0018300), and by MKE through the Fundamental R&D program for Core Technology of Materials.

†**Correspondence and requests for materials should be addressed to Kee Hoon Kim (khkim@phya.snu.ac.kr) and Kookrin Char (kchar@phya.snu.ac.kr).**

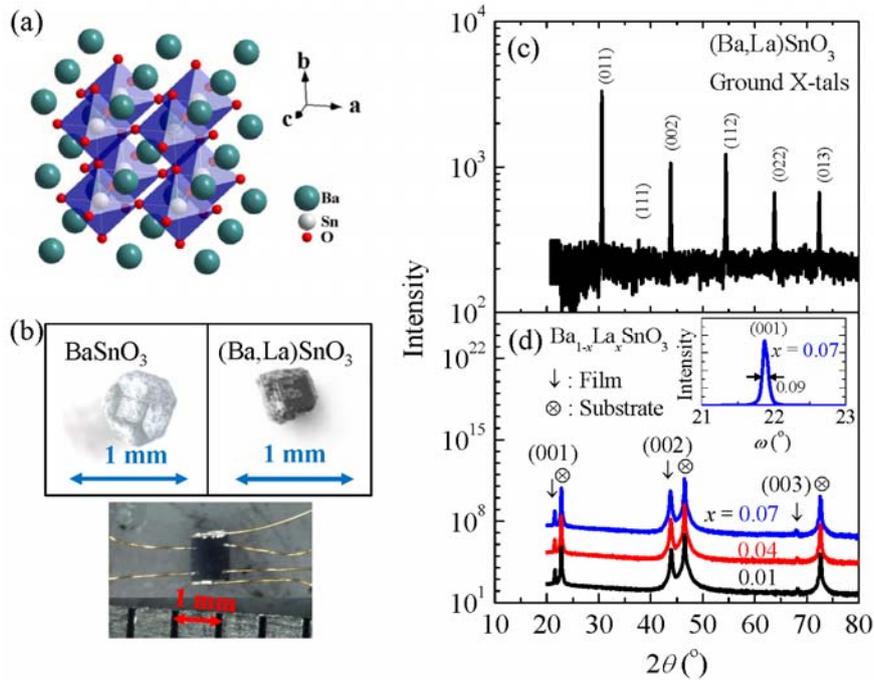

FIG. 1. (color online) (a) 2×2×2 unit cells of cubic perovskite $BaSnO_3$. (b) Optical microscope images of the flux grown $BaSnO_3$, $(Ba,La)SnO_3$ and a typical sample with electrical leads used in the transport measurements (c) X-ray $\theta$-$2\theta$ scan result of a $(Ba,La)SnO_3$ single crystal show a single phase in the cubic perovskite structure. (d) X-ray $\theta$-$2\theta$ scan results show the (00$l$) peaks in epitaxial thin films grown on $SrTiO_3$ (001) substrate. The inset figure shows that the full width at half maximum of the $\omega$-scan rocking curve is only 0.09˚.

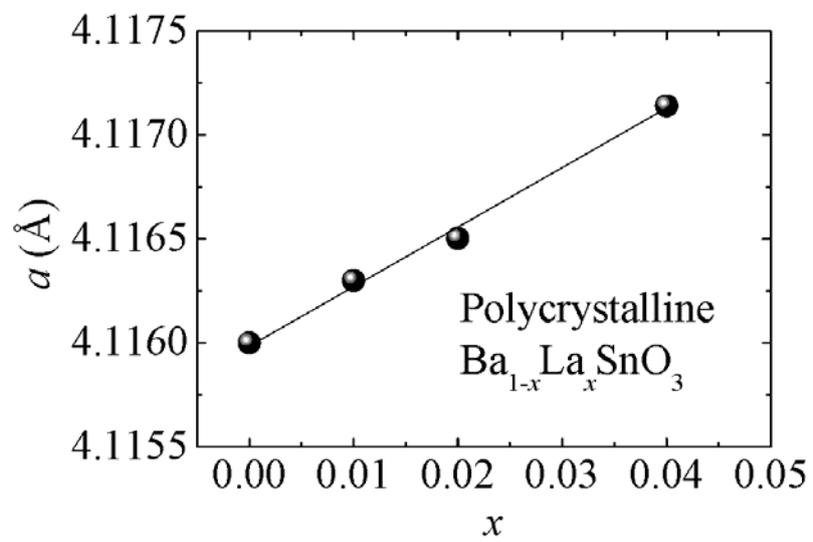

FIG. 2. (a) Cubic lattice parameter for polycrystalline $Ba_{1-x}La_xSnO_3$ ($x$ = 0, 0.01, 0.02, and 0.04) measured at RT. Line is drawn as a guide to the eye.

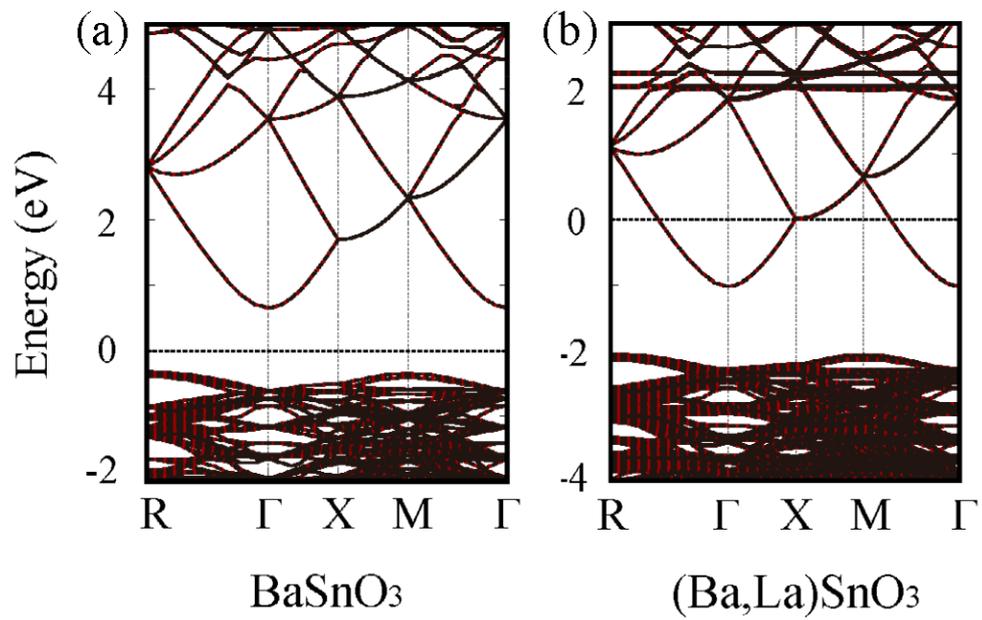

FIG 3. The band structure of (a) BaSnO$_3$ and (b) (Ba,La)SnO$_3$ obtained by first-principles calculations with 27 (3×3×3) unit cells. For (Ba,La)SnO$_3$, one of Ba$^{2+}$ ions is substituted by a La$^{3+}$ ion corresponding to the doping rate of $x$=0.037 in Ba$_{1-x}$La$_x$SnO$_3$.

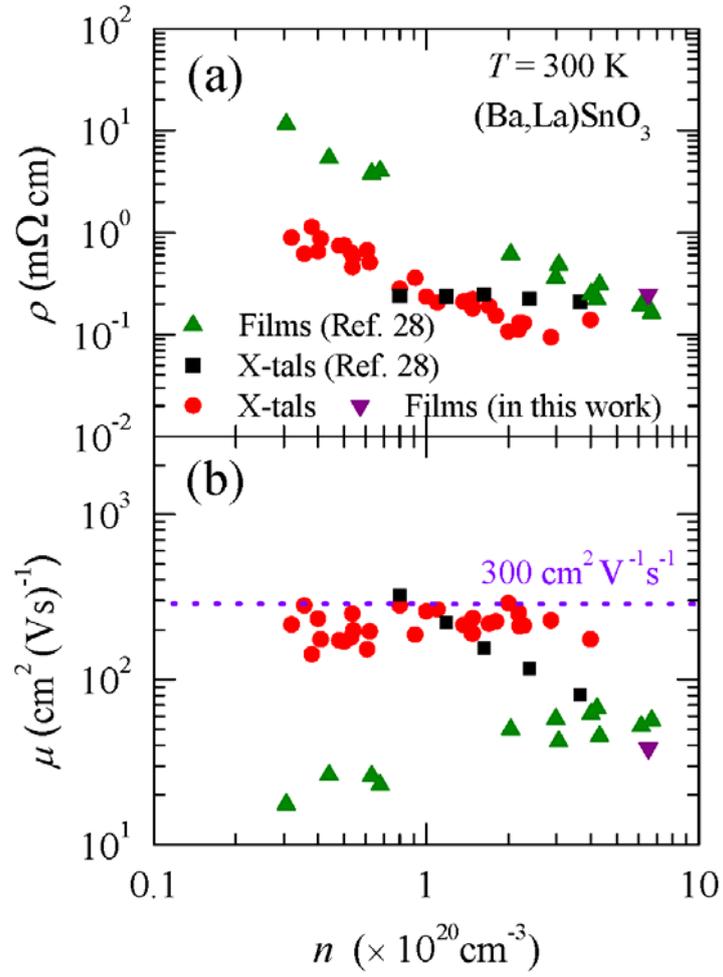

FIG. 4. (color online) (a) Resistivity ($\rho$) and (b) mobility ($\mu$) vs. carrier density ($n$) plot. closed (black) squares and (green) triangles are for the reported single crystals and films[28], respectively. Closed (red) circles are for the data of single crystals in this work.

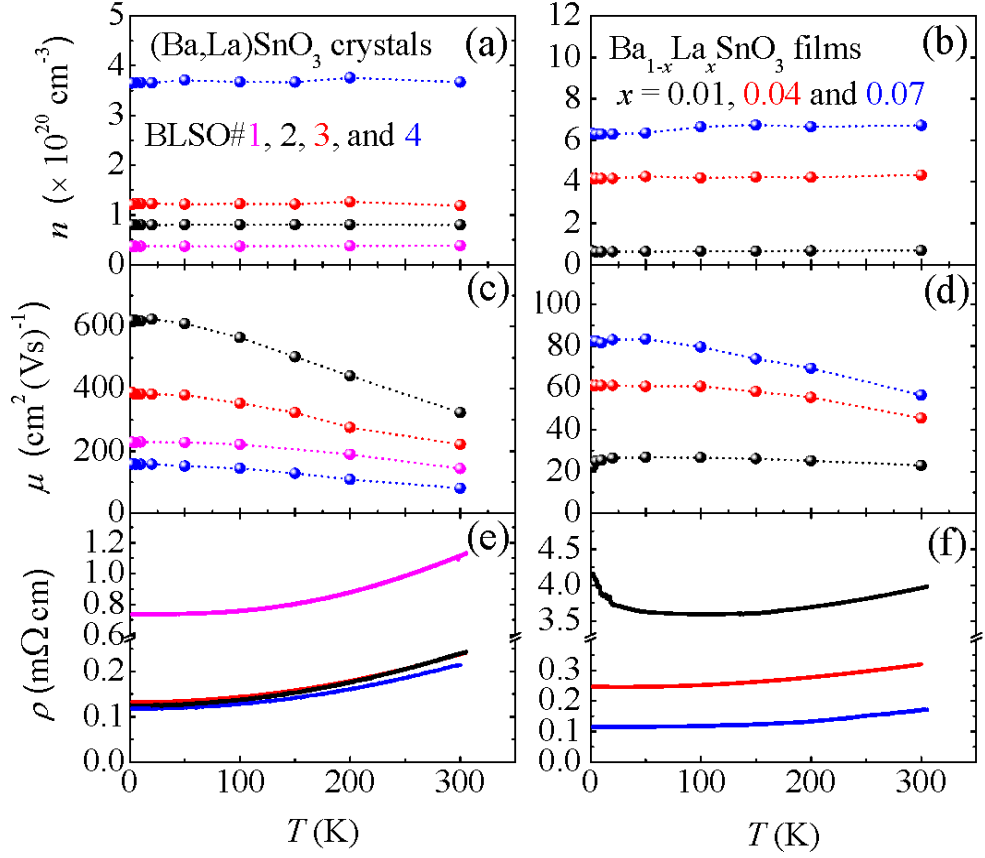

FIG. 5. (color online) Temperature-dependent carrier density $n$, resistivity $\rho$, and mobility $\mu$ are plotted for selected (Ba,La)SnO$_3$ single crystals (BLSO #1, 2, 3, and 4) (a,c,e) and thin films (b,d,f). The nominal La composition ($x$) is presented for the thin films.

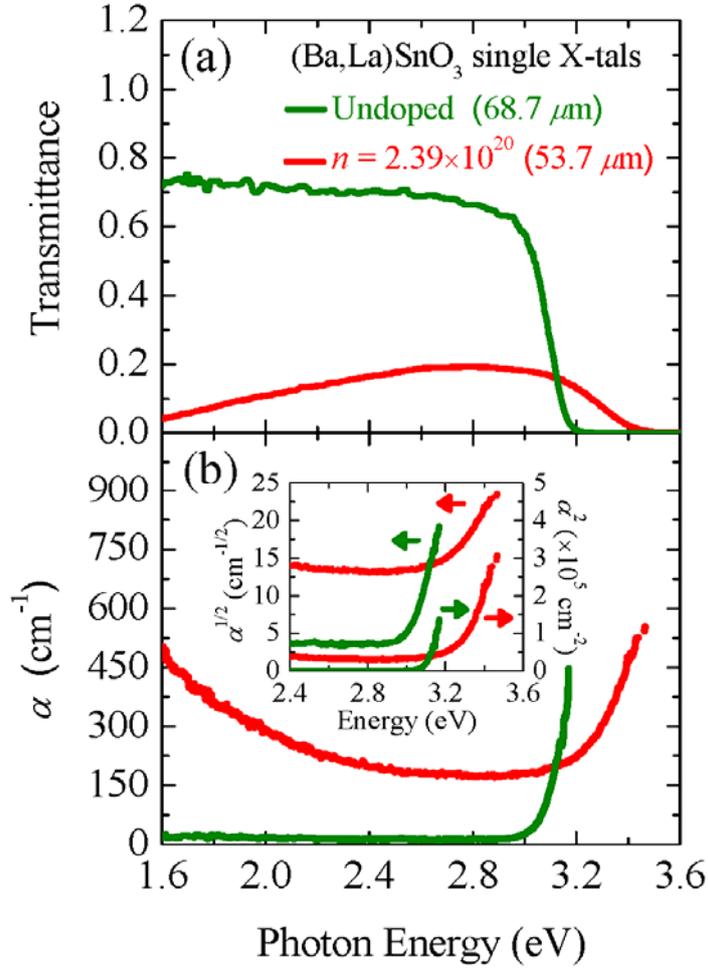

FIG. 6. (color online) (a) Transmission spectra and (b) absorption coefficient ($\alpha$) of BaSnO$_3$ (BSO) and (Ba,La)SnO$_3$ single crystals are plotted as a function of photon energy. Absorption coefficient $\alpha$ for each undoped and doped sample ($n$ = 2.39×10$^{20}$ cm$^{-3}$) was extracted by measurements of transmission spectra of the same specimen in two different thicknesses. The inset shows that the curves of $\alpha^{0.5}$ and $\alpha^2$ vs photon energy for both undoped BaSnO$_3$ and (Ba,La)SnO$_3$.

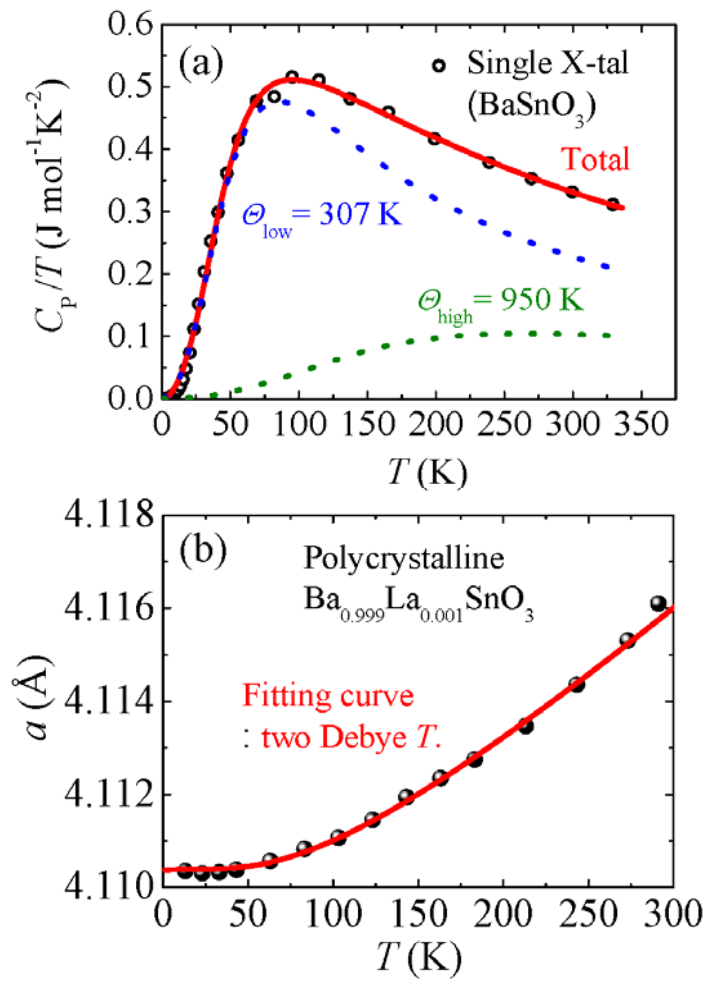

Fig. 7 (color online) (a) Open circles are the temperature dependent specific heat of BaSnO$_3$ single crystal. The dotted lines are two phonon mode contributions (blue and green) and the solid line refers to their sum (red). (b) Thermally induced cubic lattice parameter expansion of polycrystalline Ba$_{0.999}$La$_{0.001}$SnO$_3$. The solid (red) line is the thermal expansion fitting based on the Debye model with the two Debye temperatures.